\documentstyle[psfig]{l-aa}

\begin{document}

\thesaurus {06 (04.03.1, 08.02.5)}

\title{UBV(RI)$_{\rm C}$ photometric comparison sequences for symbiotic stars}

\author{
       Arne Henden\inst{1}
\and   Ulisse Munari\inst{2}
       }
\offprints{U.Munari}

\institute {
Universities Space Research Association/U. S. Naval Observatory
Flagstaff Station, P. O. Box 1149, Flagstaff AZ 86002-1149, USA
\and
Osservatorio Astronomico di Padova, Sede di Asiago, 
I-36012 Asiago (VI), Italy
}
\date{Received date..............; accepted date................}

\maketitle

\markboth{A.Henden and U.Munari: UBV(RI)$_{\rm C}$ photometric comparison sequences for 
symbiotic stars}{A.Henden and U.Munari: UBV(RI)$_{\rm C}$ photometric comparison 
sequences for symbiotic stars}

\begin{abstract}
We present accurate UBV(RI)$_{\rm C}$ photometric sequences around 20 symbiotic stars.
The sequences extend over wide brightness and color ranges, and are suited
to cover quiescence as well as outburst phases. The sequences are intended
to assist both present time photometry as well as measurement of
photographic plates from historical archives.
\end{abstract}
\maketitle

\section{Introduction}

Symbiotic stars are binary systems composed of a cool giant and a hot,
luminous white dwarf. They show variability over any time scale from minutes
(flickering) to several decades (outbursts of symbiotic novae), with
phenomena related to the orbital motion having periodicities generally
between 1 and 4 years (or a few decades in the systems harbouring a
Mira variable, $\sim$20\% of all known symbiotic stars).

Such long time scales tend to discourage stand--alone photometric campaigns
from a single Observatory (which would require observational programs
running up to $\sim$10 years in order to derive - for example - a firm orbital
period). Most of the current photometric investigations of symbiotic stars
therefore try to assemble as much as possible data from the widest set of
current and archival sources. A template example is the recent
reconstruction of the 1890-1996 lightcurve and orbital period determination
for YY~Her by Munari et al. (1997).

The data so collected are generally very heterogeneous in nature, with large
differences caused for example by ($a$) the non-standard photometric bands,
($b$) the adopted comparison sequences and standard stars, ($c$) lack of
adherence to and transformation into a system of general use, like the
UBV(RI)$_{\rm C}$, and ($d$) telescope focal length or pixel scale
that causes blending with images of nearby field stars. These
differences generally may introduce such a large scatter in the data that
all but the strongest details are washed out.

The establishment of suitable and accurate photometric comparison sequences
covering a wide range in magnitude and colors should alleviate considerably
some of the above problems, and could encourage small observatories and/or
occasional observers to obtain new data as well as to encourage 
those with access to plate archives to search for valuable historical data.

To this aim we present here suitable, UBV(RI)$_{\rm C}$ comparison sequences
for 20 symbiotic stars (all but a few accessible from both hemispheres. See
Table~1 for a list of the program stars). The sequences are basically
intended to allow a general observer to capture on a single CCD frame or to
have in the same eyepiece field of view when inspecting archival
photographic plates: ($a$) enough stars to cover the whole range of known or
expected variability for the given symbiotic star, ($b$) stars of enough
different colors to be able to calibrate the instrumental color equations
and therefore reduce to the standard UBV(RI)$_{\rm C}$ system the collected
data, and ($c$) stars well separated from surrounding ones to avoid blending
at all but the shortest telescope focal lengths.

\begin{table*}
\caption[]{List of program symbiotic stars. The coordinates for the symbiotic stars
are from our observations (equinox J2000.0, mean epoch 1999.5). 
The $e_\alpha$ and $e_\delta$ columns list the errors in milliarcsec
for right ascension and declination, respectively. The last two
columns list the coordinates of the field centers in Figures~1 and 2.}
\centerline{\psfig{file=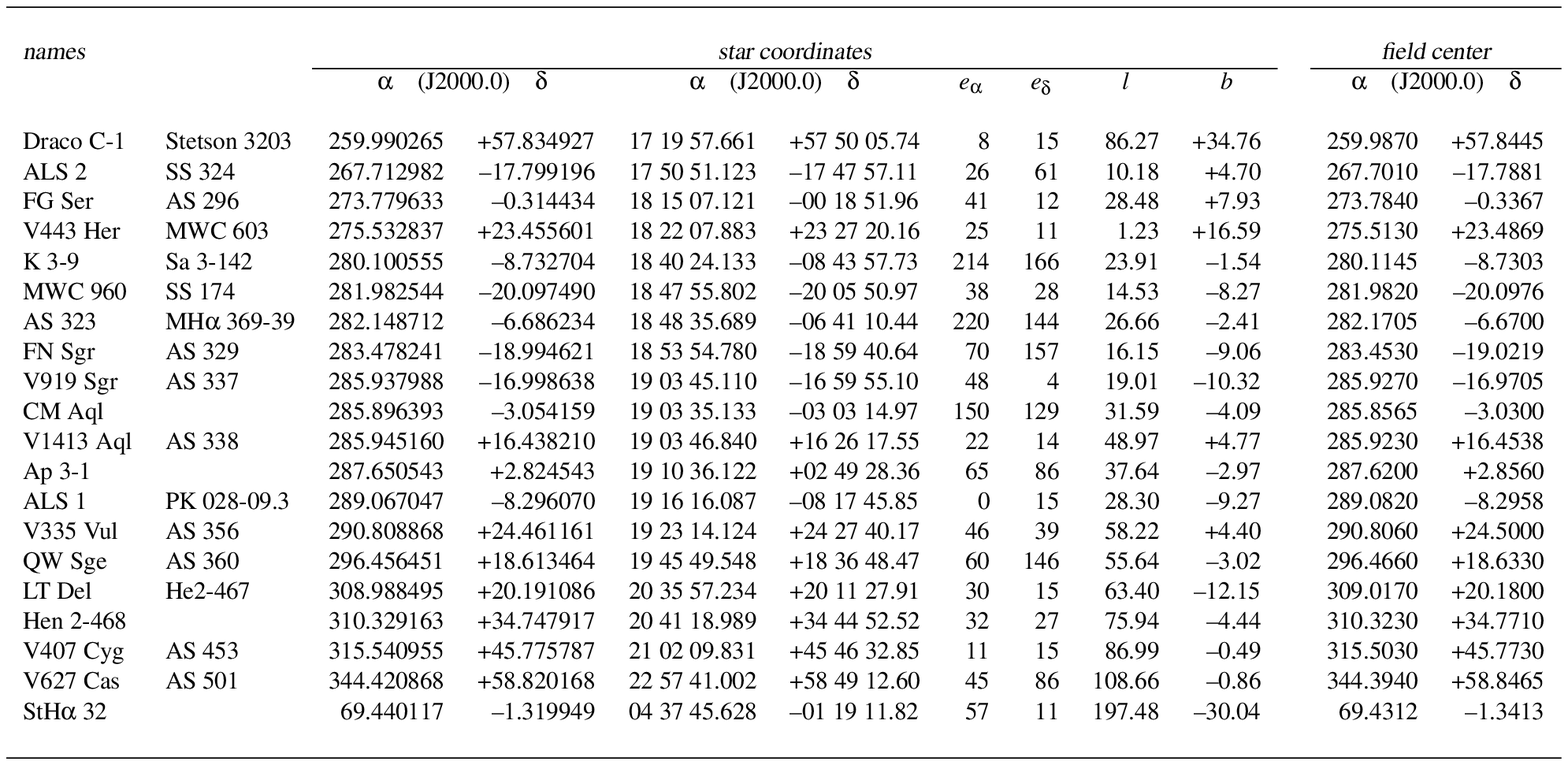,width=18cm}}
\end{table*}

\begin{figure*}[h!]
\centerline{\psfig{file=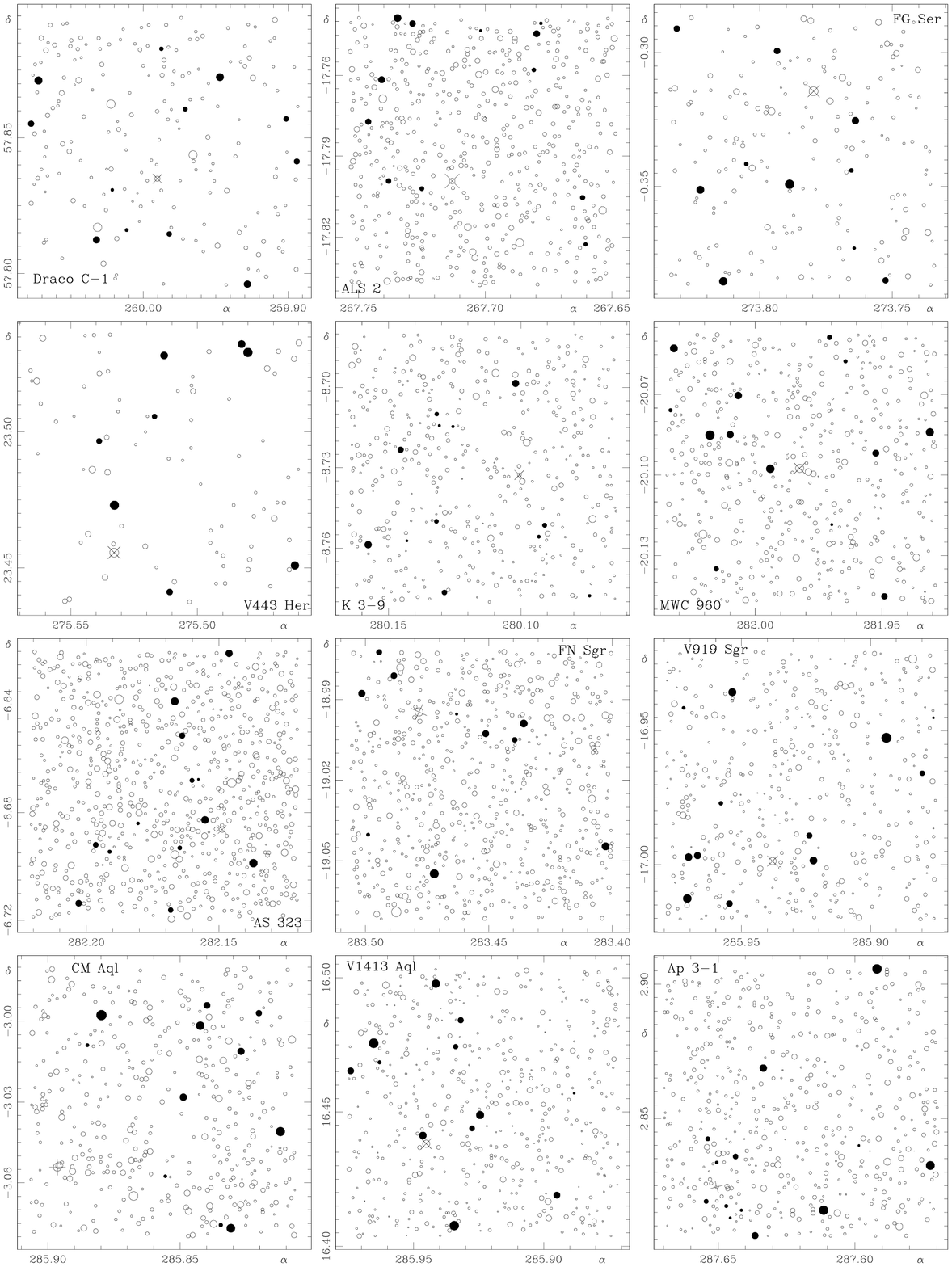,width=17.5cm}}
\caption[]{Finding charts for the UBV(RI)$_{\rm C}$ comparison
photometric sequence around the program symbiotic stars. The fields are
in the same order as in Table~1. North is up and East to the left,
with an imaged field of view of 5.16$\times$5.16 arcmin and
a 5.4$\times$5.4 coordinate grid (see bottom-right panel of Figure~2).
Stars are plotted as open circles of diameter proportional to the
brightness in the $V$ band. The stars making up the photometric sequence
(see Table~2) are plotted as filled circles.}
\end{figure*}

\section{Observations}

All observations were made with the 1.0-m Ritchey-Chr\'etien telescope of
the U. S. Naval Observatory, Flagstaff Station. A Tektronix/SITe 1024x1024
thinned, backside--illuminated CCD was used, along with Johnson UBV and
Kron--Cousins RI filters. Images were processed using IRAF, with nightly
median sky flats and bias frames.  Aperture photometry was performed with
routines similar to those in DAOPHOT (Stetson 1987). Astrometry was
performed using SLALIB (Wallace 1994) linear plate transformation routines
in conjunction with the USNO--A2.0 reference catalog. Errors in coordinates
were typically under 0.1 arcsec in both coordinates, referred to the mean
coordinate zero point of the reference stars in each field.

The telescope scale is 0.6763 arcsec/pixel, with a total field of view of
around 11.4x11.4 arcmin.  Typical seeing was $\sim$2 arcsec. A 9 arcsec
extraction aperture with concentric sky annulus was commonly used.

The reported photometry only uses data collected on photometric nights
(transformation errors under 0.02mag).  For each such night, symbiotic field
observations were interspersed with observations of Landolt (1983, 1992)
standard fields, selected for wide color and airmass range.  The mean
transformation coefficients (cf. Henden \& Kaitchuck
1990, eqns. 2.9ff) are:

\begin{eqnarray}
  V:& -0.020& \pm 0.007 \\
B-V:& ~0.949& \pm 0.007 \\
U-B:& ~1.072& \pm 0.018 \\
V-R:& ~1.017& \pm 0.005 \\
R-I:& ~0.971& \pm 0.013 
\end{eqnarray}

Second order extinction was negligible except for {\sl B--V}, where a
coefficient of --0.03 was used.

The symbiotic field photometry was usually performed as the field transited. 
In a few rare cases, observations were made at higher airmass to the West. 
In these cases, care was taken to obtain extinction measures at equivalent
or higher airmass.  Each field was observed on at least three nights.  Since
all primary standards used the same aperture as the secondary standards
being established, and the apertures were large, no aperture corrections
were necessary.

\section{The photometric sequences}

Between 10 and 15 stars around each symbiotic star have been selected to
form the comparison sequences, given in Table~2. The sequences have been
selected and ordered on the basis of the $B$ magnitude. The $B$ magnitude is
reproducible by most filter-equipped CCD cameras, it is the closest one to
the $m_{pg}$ band of the historical photographic observations and the $B$
band is particularly well suited to investigate the variability of
symbiotic stars (see next section).

The range in magnitude of the sequences is large enough to cover both
outburst and quiescence phases (eclipses included) of each symbiotic star.
The comparison sequences are tighter around the usual brightness of the
symbiotic stars and become looser away from it. In most cases the sequences
extend to much fainter magnitudes (down to $B \geq 19$ mag or $B \geq 20$
mag) than reached by the respective symbiotic stars because they could be of
interest to other observational projects as well as in assisting in the
calibration of sky survey projects on photographic plates.

For 9 objects (Draco C-1, ALS~2, K~3-9, V919~Sgr, Ap~3-1, V335~Vul,
Hen~3-468, V627~Cas and StH$\alpha$~32) the symbiotic star and the
comparison sequence both lie inside a 5.16$\times$5.16 arcmin field (see
Figure~1), which match in dimension the Allen (1984) finding charts. For the
remaining 11 program objects, comparison stars bright enough to cover the
outburst phases had to be found at greater distance from the symbiotic star.
They are given in Table~2 at the end of each sequence, separated by an empty
line from the other comparison stars, and are plotted on the less deep and
wider finding charts of Figure~3.

The stars included in the comparison sequences have been checked on at least
three different nights for variability (see column $N$ of Table~2). We
cannot rule out beyond doubt that some of them are indeed variable (they
could be eclipsing systems observed outside eclipse, for example), but the
fairly good  

\begin{figure*}[!h]
\centerline{\psfig{file=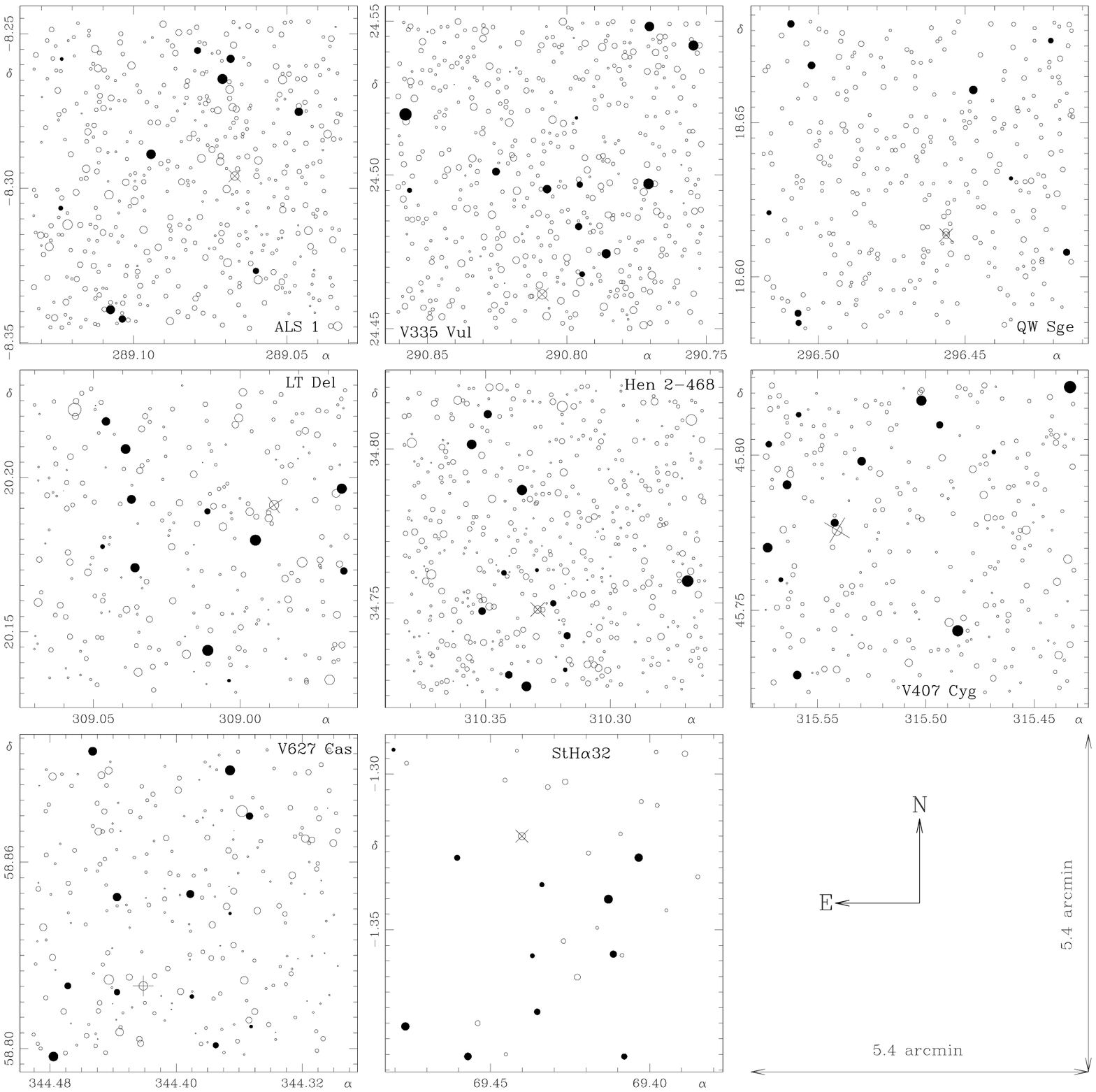,width=17.5cm}}
\caption[]{Same as Figure~1.}
\end{figure*}

\begin{figure*}[!h]
\centerline{\psfig{file=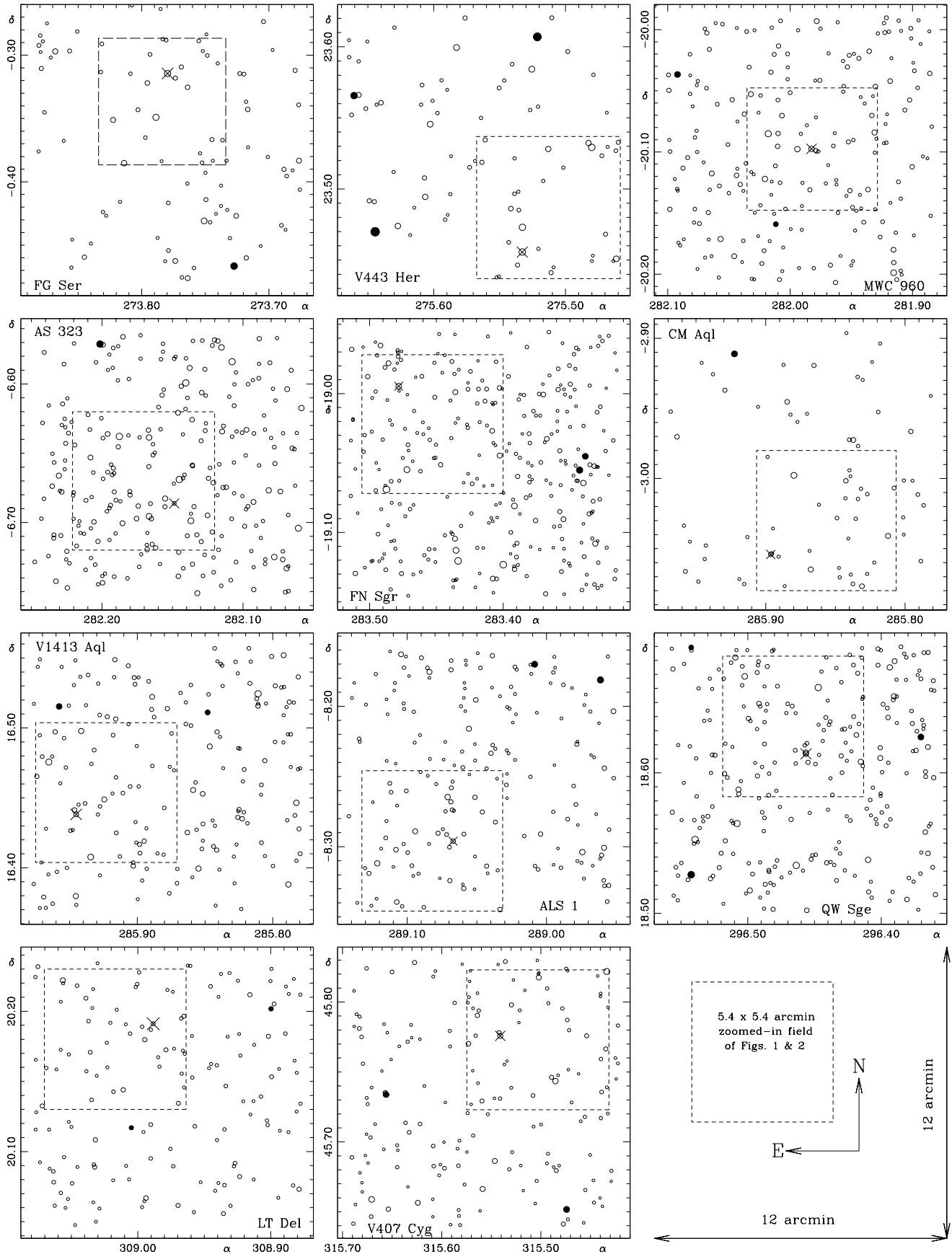,width=17.5cm}}
\caption[]{Comparison stars bright enough to cover the outburst phases
lay outside the fields of Figures~1 and 2 for eleven symbiotic stars.
They are identified in these wider finding charts, where the portion 
plotted in greater detail in Figures~1 and 2 is outlined by
a dashed square. The symbols are the same as in Figures~1 and 2, with
an imaged field of view of about 11.4$\times$11.4 arcmin and a
12$\times$12 arcmin coordinate grid. To avoid overcrowding, the limiting
magnitude is much brighter ($V \sim$16 mag) than in Figures~1 and 2.}
\end{figure*}

\clearpage

\noindent
agreement (at a few millimag level) of their magnitudes as measured on
different nights over some months gives some confidence in their use. 
Finally, to avoid problems of blending with nearby stars on plates or CCD
images from short focus telescopes, the comparison stars have been selected
so as to avoid those with close companions.

\section{The type of variability in symbiotic stars}

As a guideline for observers not familiar with the symbiotic stars, a few
simplified notes may be of interest concerning the types of variability one
may expect from the latter and the best way to observe them. We will limit
the discussion to the photometric bands of the UBV(RI)$_{\rm C}$ system.

The variability ascribed to the {\sl cool giant} is best observed at longer
wavelengths (i.e. the $I$ band. The $R$ band is affected by the usually
very strong H$\alpha$ emission), where the contamination from the white
dwarf companion and the circumstellar material become less important.
Basically, two types of variability of the cool giant may be observed:\\
  \underline{\sl intrinsic}, like the pulsations of a Mira variable (about
  $\sim$ 20\% of the known symbiotics harbor a Mira). The amplitude of
  variability generally decreases toward longer wavelengths. Popular examples
  are R~Aqr (pulsation period of 386 days, minima as faint as $V$=12, maxima
  as bright as $V$=5 mag) or UV Aur (pulsation period of 395 days, minima as
  faint as $V$=11, maxima as bright as $V$=7.5 mag);\\
  \underline{\sl ellipsoidal}, when the cool giant fills its Roche lobe.
  Due to the orbital motion the area of the Roche lobe projected onto
  the sky varies continuously, with two maxima (when the binary system is
  seen at quadrature) and two minima (when the cool giant passes at superior
  or inferior conjunctions) per orbital cycle. Because the reason for
  variability is a geometrical one, the amplitude of variability is not strongly
  dependent upon wavelength.  Popular examples of symbiotic stars showing
  ellipsoidal distortion of their lightcurve are T~CrB (orbital period 227
  days, amplitude $\bigtriangleup m$= 0.3 mag) and BD--21.3873 (orbital
  period 285 days, amplitude $\bigtriangleup m$= 0.2 mag).

The variability ascribed to the {\sl hot white dwarf} companion to the cool
giant is best observed at shorter wavelengths. There are several type of
manifestations, among which :\\
  \underline{\sl outbursts}, with amplitudes $\bigtriangleup B  = 2 -
  5$ mag and duration from half a year to many decades. The amplitude,
  duration and lightcurve shape are usually unpredictable.  The same system
  might show completely different type of outbursts one after the other. 
  For example, QW~Sge had an outburst extending from July 1962 to March 1972
  characterized by a rapid rise and a linear and smooth decline, followed by
  another one from May 1982 to September 1989 showing a complex lightcurve
  with more than one maximum and deep minima in between;\\
  \underline{\sl reflection effect}, when the hard radiation field of the
  hot and luminous white dwarf (radiating mainly in the X-ray and far
  ultraviolet domains) illuminates and heats up the facing side of the cool
  giant (which reprocesses to the optical domain the energy received by the
  white dwarf). The heated side of the cool giant is therefore a bit
  brighter and bluer than the opposite one (which is not illuminated by the
  white dwarf radiation field). During an orbital period the heated side comes
  and goes from view, causing a sinusoidal lightcurve.  The effect is strongly
  wavelength dependent, being maximum in the $U$ band and undetectable in
  $R$ and $I$ bands.  The amplitude may be fairly large, as in LT~Del where
  $\bigtriangleup U$=1.6, $\bigtriangleup B$=0.5 and $\bigtriangleup V$= 0.2
  mag. It should be observable in the majority of symbiotic stars (more and
  more easily as the white dwarf gets hotter and the orbital inclination
  increases) and it is a powerful way to measure orbital periods;\\
  \underline{\sl eclipses} of the white dwarf by the cool giant.  In
  quiescence the eclipses generally escape detection by optical photometry
  because the white dwarf is radiating mostly at shorter wavelengths (X-rays
  and far ultraviolet).  During the outbursts the white dwarf emission
  shifts to longer wavelengths and becomes conspicuous in the optical, thus
  allowing the eclipses to be detected if the orbital inclination is
  sufficiently high.  Classical examples of symbiotic stars for which the
  eclipses passed undetected in quiescence and instead became outstanding
  features of the outburst lightcurve are FG~Ser and V1413~Aql.  Because
  the eclipsing body is cool and the eclipsed one is hot, the visibility of
  eclipses increases toward shorter wavelengths (for example for FG Ser in
  outburst it was $\bigtriangleup V$=1.4, $\bigtriangleup B$=1.9 and
  $\bigtriangleup U$=2.3 mag);\\
  \underline{\sl re-processing} by the circumstellar nebula of the energy
  radiated by the white dwarf.  Sometimes there is so much circumstellar gas
  ionized by the radiation field of the white dwarf that its brightness
  completely overwhelms that of the binary system, as it is for the popular
  cases of V1016~Cyg and V852~Cen (the {\sl Southern Crab}). Both these symbiotic
  binaries harbor a Mira variable, whose variability however does not at all
  affect the optical photometry because of the immensely brighter
  circumstellar ionized gas. When the white dwarf becomes progressively
  cooler and dimmer, the amount of ionizing photons that it releases goes
  down, and the ionized fraction of the circumstellar nebula decreases and
  consequently its brightness (the scenario is valid for radiation bounded
  nebulae). This is the case for HM~Sge that over the last 25 years has
  gradually become fainter by $\bigtriangleup V$ = 0.031 and
  $\bigtriangleup B$ = 0.086 mag~yr$^{-1}$.
  
\section{Notes on individual symbiotic stars}

A few individual notes follow on the photometric behavior of the program
symbiotic stars, to the aim of assisting the interested reader in planning
an observing strategy. An inspiring reading would also be the collected
history of symbiotic stars assembled by Kenyon (1986).

While calibrating the photometric comparison sequences for this paper we
have also collected data on the program symbiotic stars. These UBV(RI)$_{\rm
C}$ data will be presented and discussed elsewhere together with similar
data for more than another 100 symbiotic stars observed from ESO and Asiago.
To the reader's benefit we report in this section mean $B$ and {\sl B-V}
values for 1999 from the UBV(RI)$_{\rm C}$ survey (hereafter indicated as
MHZ: Munari, Henden and Zwitter, in preparation).

\underline{\sl Draco C-1}. This carbon symbiotic star belongs to the Draco
dwarf galaxy (Aaronson et al. 1982). Infrared photometry by 

\begin{table*}
\caption[]{The comparison sequences around the 20 program symbiotic stars. Positions
for J2000 equinox and a mean epoch 1999.5 are given (errors in arcsec are
derived from different exposures in different bands), together with
magnitudes and colors (errors in magnitudes). The stars in each sequence are ordered 
according to fainter $B$ magnitudes. $N$ is the number of observing nights.
The sequences are given in the same order as in Table~1 and Figures~1 and 2.
The comparison stars laying outside the field of view of Figures~1 and 2
and plotted in Figure~3 are given at the bottom of each sequence, separated by
an empty line.}
\centerline{\psfig{file=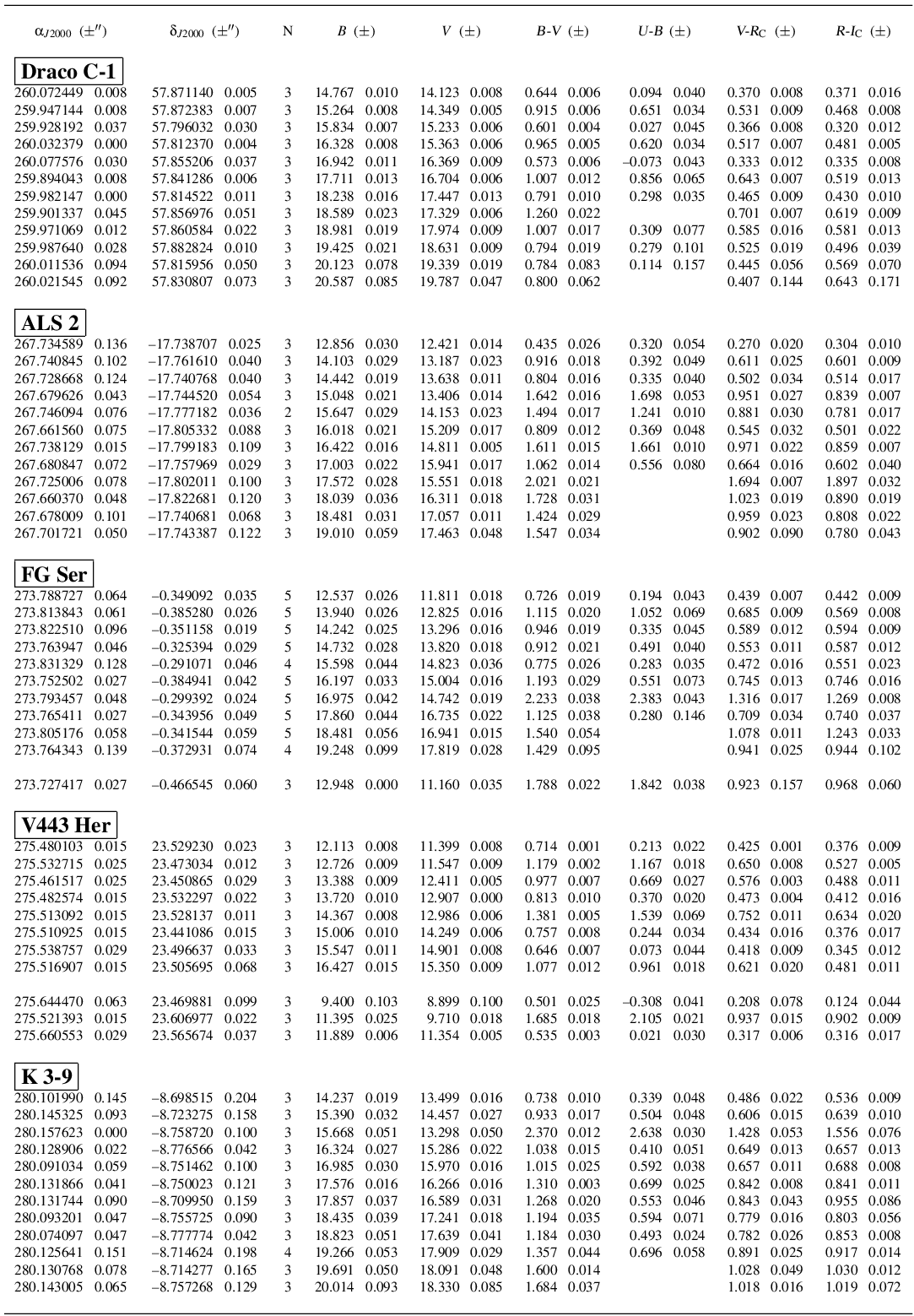,height=23.5cm}}
\end{table*}

\setcounter{table}{1}
\begin{table*}
\caption[]{({\sl continues})}
\centerline{\psfig{file=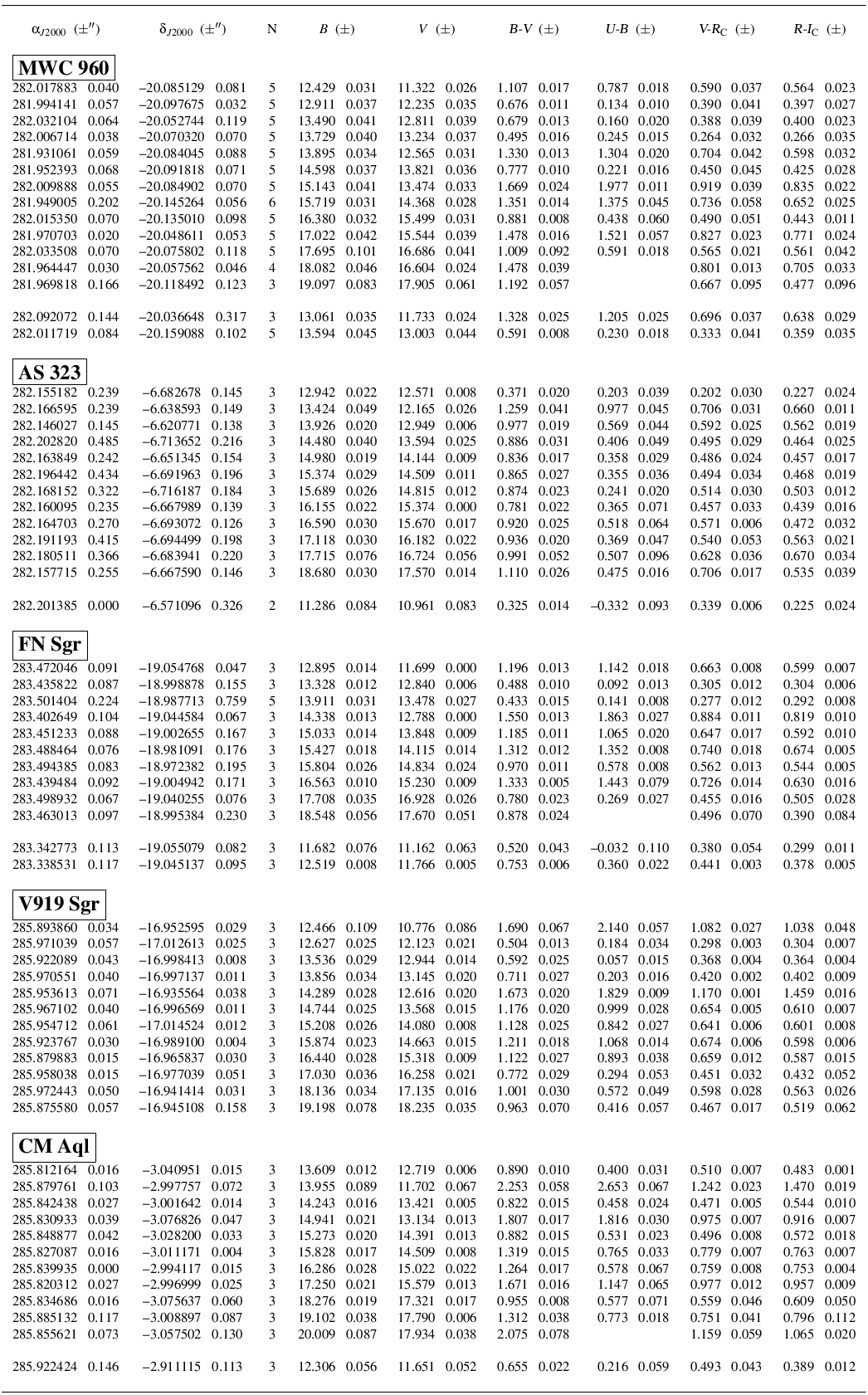,height=24.5cm}}
\end{table*}

\setcounter{table}{1}
\begin{table*}
\caption[]{({\sl continues})}
\centerline{\psfig{file=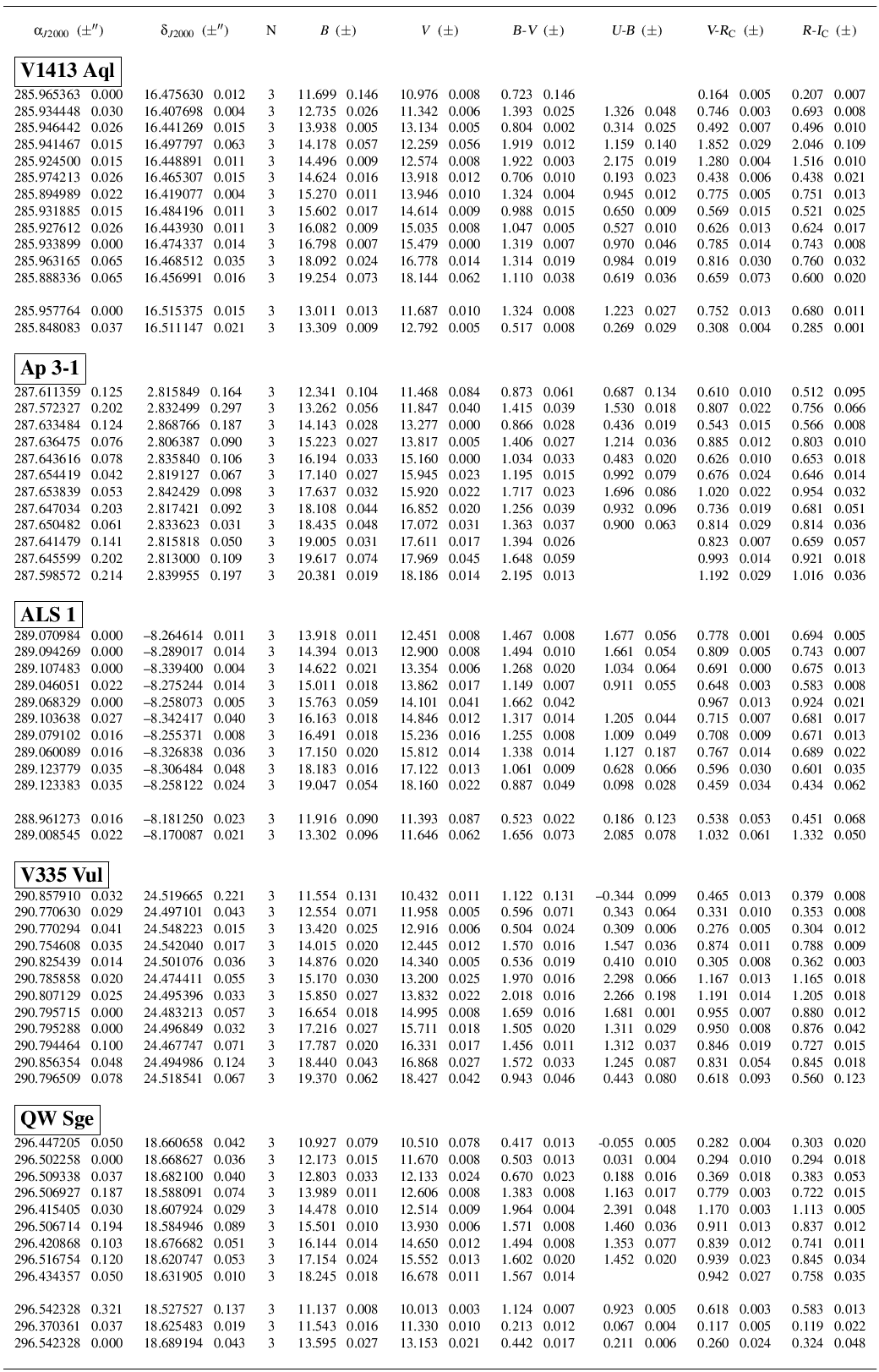,height=24.5cm}}
\end{table*}

\setcounter{table}{1}
\begin{table*}
\caption[]{({\sl continues})}
\centerline{\psfig{file=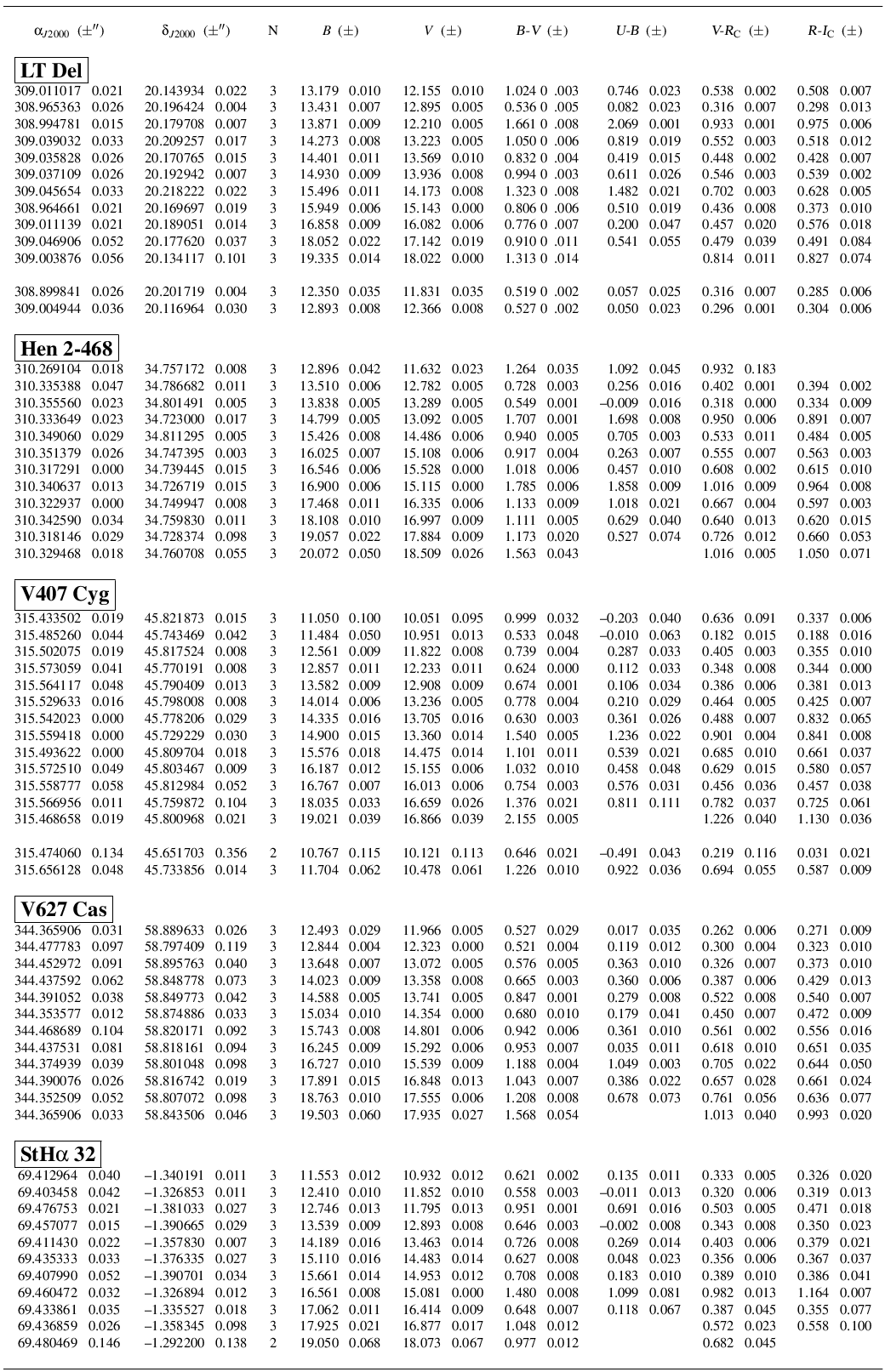,height=24.5cm}}
\end{table*}

\noindent
Munari (1991a) proves C-1 to be at the tip of the Draco AGB with very blue
IR colors for a carbon star, probably caused bythe low metal content of the
parent galaxy (Munari 1991b). No outburst has been so far recorded and the
orbital period is unknown. {\sl BVI} photometry by Munari (1991c) seems to
support a variability of the carbon giant with a period of $\sim$ 55 days.
If confirmed, this would be among the shortest period known for carbon
pulsating variables (cf. Claussen et al.  1987). MHZ report $B$=18.6 and
{\sl B--V}=+1.5 mag.
 
\underline{\sl ALS 2}. Its symbiotic nature has been discovered by Acker et
al. (1988). MHZ lists $B$=16.2 and {\sl B--V}=+1.9 mag.  The orbital
period, type of variability and presence of historical outburst are unknown.

\underline{\sl FG Ser}. After the 1988--1994 outburst when it rose to
$B$=10.4 and {\sl B--V}=+1.1, it is now back toward the quiescent $B$=13.8
and {\sl B--V}=+2.0 mag values. MHZ list $B$=13.5 and {\sl B--V}=+1.7 mag.
Munari et al. (1992b) discovered eclipses during the outburst phase (of
amplitude $\bigtriangleup V$=1.4, $\bigtriangleup B$=1.9, $\bigtriangleup
U$=2.3 mag and 120 days between first and fourth contact). From three
consecutive mimina Munari et al. (1995) derived the following ephemeris
\begin{equation}
T(min) = 2448492 (\pm 4) \ + \ 658 (\pm 4) \times E
\end{equation}
where 658 is the period in days, as usual.
From 250 archive blue plates covering the period 1949-1987, Kurockin (1993)
found a large and sinusoidal variability at quiescence ($\bigtriangleup
B$=1.5 mag), following the ephemeris
\begin{equation}
T(min) = 2446591 \ + \ 630  \times E
\end{equation}
which could be interpreted in terms of a reflection effect. The difference
between the two periods (both should trace the orbital period) has to be
investigated. It should also be noted that the cool component does not show
intrinsic or ellipsoidal variability in excess of 0.1 mag. The eclipses have
not yet been searched for during quiescence. Their detection and monitoring
would be of interest to measure the size, temperature and luminosity
of the white dwarf now that it is returning back to quiescence dimensions.

\underline{\sl V443 Her}. No outburst has ever been recorded from this
fairly bright symbiotic star. Its behavior in quiescence has been investigated by
Kolotilov et al. (1995) who found a lightcurve dominated by a reflection effect 
of $\bigtriangleup U$=0.9, $\bigtriangleup B$=0.4 and $\bigtriangleup V$= 0.1 mag
amplitude. The minima follow the ephemeris
\begin{equation}
T(min) = 2443660 (\pm 30) \ + \ 594 (\pm 3)  \times E
\end{equation}
There seems to be another periodicity of no easy interpretation at 430 days.
The mean values in quiescence are $B$=12.5 and {\sl B--V}=+1.0. Limited
infrared observations by Kolotilov et al. (1998) seems to argue against
variability of the cool giant or an ellipsoidal distortion of it.

\underline{\sl K 3-9}. Originally classified among the planetary nebulae,
its symbiotic star nature has been discovered by Acker et al. (1983).
According to Ivison and Seaquist (1995) K 3-9 is among the brightest
symbiotic radio sources, and could harbor a Mira variable and a WD locked in
a permanent outburst state. A thick dust cocoon should encircle the binary
system, and a huge external ionized nebular material completely dominates
the optical spectra. The photometric properties, history and orbital period
are unknown. MHZ report $B$=18.3 and {\sl B--V}=+1.3.

\underline{\sl MWC 960}. This is a bright symbiotic star neglected by the
observers. Munari et al. (1992a) report $B$=13.6 and {\sl B--V}=+1.5 and MHZ
list $B$=13.8 and {\sl B--V}=+1.6. The photometric properties, history and
orbital period are unknown.

\underline{\sl AS 323}. Another object originally classified as a planetary
nebula which later turned out to be a symbiotic star (Sabbadin 1986, Acker et
al. 1983). MHZ report $B$=15.2 and {\sl B--V}=+1.0. The photometric
properties, history and orbital period are unknown.

\underline{\sl FN Sgr}. Another bright symbiotic star that has been
overlooked by most observers even though reports of large variability date
back to Ross (1926). Outbursts have been recorded in 1924-1926 and
1936-1941. The brightness in quiescence seems to vary by a large amplitude
($\bigtriangleup m \sim$2 mag) with possible periodicities between 1 and 3
years (cf. Kenyon 1986 and references therein). Amateur visual observations
over the last few years show a pattern reminiscent of an eclipsing binary
following the ephemeris (Munari et al., in preparation):
\begin{equation}
T(min) = 2451410 (\pm 15) \ + \ 1120 (\pm 20)  \times E
\end{equation}
The mean brightness is $V \sim$13.5 in eclipse and $V \sim$11.0 outside.
Next minimum is scheduled for mid September 2002. MHZ list $B$=12.7 and {\sl
B--V}=+0.7.

\underline{\sl V919 Sgr}. Another bright object ignored by observers.
According to literature review and new observations by Ivison et al. (1993),
V919 Sgr varies between $B$=12 and $B \geq$ 14.2 mag. Its cool giant is
definitively variable in the infrared by at least $\bigtriangleup K$=0.7
mag. The announcement of an outburst was made in 1991, but it is not yet
proven it actually occurred (see Ivison et al. 1993). MHZ report $B$=14.2
and {\sl B--V}=+1.2. The past photometric history and orbital period are
unknown.

\underline{\sl CM Aql}. Another relatively bright symbiotic star that has
been overlooked by most observers. Its range of variability extend from
$B$=13.0 to $B$=16.5. Outbursts have been reported for 1914, 1925, 1934,
1950. CM Aql also attracted some attention in late 1992 when from the usual
$V=13.2$ it rose for a short period to $V\sim 12$ mag. The orbital period is
unknown. MHZ report $B$=14.6 and {\sl B--V}=+1.3

\underline{\sl V1413 Aql}. The star erupted into a symbiotic nova in late
1981, and has not yet returned to quiescence conditions. According to
Munari (1992), V1413 Aql presented in quiescence 
one of the largest known reflection effects ($B$ varying between 16.5 and
14.0 mag). When the star erupted into outburst, deep eclipses appeared
perfectly in phase with the minima of the reflection effect according the
ephemeris
\begin{equation}
T(min) = 2446650 (\pm 15) \ + \ 434.2 (\pm 0.2)  \times E
\end{equation}
At outburst maximum the star peaked at $B$=11.2 and {\sl B--V}=+0.7
(compared to $B$=15.5 and {\sl B--V}=+1.5 in quiescence). The decline has
been very slow but smooth until late 1992 when V1413 Aql went back on the
rise and returned to peak brightness ($V$=10.5) by summer of 1995 and
started to decline again in a very smooth way (Munari 1996). The
eclipses have always been visible during the whole outburst phases since
1982, and at minimum the star shines at $V\sim$15.0. Outside eclipses the
star is currently at $V$=13.1 and {\sl B--V}=+0.9. If the present rate of
decline of $\bigtriangleup V$=0.56 mag yr$^{-1}$ will be maintained in the
future, the star should return to quiescence brightness by late 2002. A
detailed multi-band monitoring of successive eclipses would be of great
interest to model the radius, temperature and luminosity of the outbursting
component while it is returning back to quiescence conditions.

\underline{\sl Ap 3-1}. Another example of an object originally classified 
as a planetary nebula, and which later turned out to be a symbiotic star (Allen 1984).
Its photometric properties are unknown. MHZ list $B$=19.1 and
{\sl B--V}=+2.1

\underline{\sl ALS 1}. Its symbiotic nature has been discovered by Acker et
al. (1988), who report $V$=14.8 mag. MHZ lists $V$=13.5 and {\sl
B--V}=+1.4 mag. The photometric properties, history and orbital period are
unknown.

\underline{\sl V335 Vul}. A symbiotic nature for this carbon star has been
suggested only recently (Munari et al. 1999a). The star colors are very
red, with MHZ giving $V$=12.9 mag and {\sl B-V}=+5.1 for quiescence.
Munari et al. (1999b) caught the star on the rising branch of an
apparent outburst, with $V$=11.3 and {\sl B--V}=+3.1 and a ten-fold increase
in the intensity of emission lines. The orbital period is unknown.
According to Dahlmark (1993) the carbon star is a Mira variable of
10.5 $< V <$ 13.2 range, and maxima given by the ephemeris
\begin{equation}
T(min) = 2446740 \ + \ 342 (\pm 0.2)  \times E
\end{equation}
The possible outburst reported by Munari et al. (1999b) for the end of 1999
happened at the time of maximum brightness for the Mira variable (O.Pejcha
and P.Sobota, private communication). The puzzling coincidence of the two
events should be studied in more details.

\underline{\sl QW Sge}. Examining 438 archive blue plates covering the
period 1960-1992, Kurockin (1993) has discovered two outbursts: one extending
from July 1962 to March 1972 with $B$=11.5 at maximum, the other from May
1982 to September 1989 with a much more complex lightcurve and a peak
brightness $B$=12.0. 
Outside outburst phases the star is since first observations in 1898 at
$B\sim 13.1\div 13.3$.  MHZ lists $B$=13.2 and {\sl B--V}=+0.81 mag. QW Sge
has an optical companion 3.5 arcsec to the north, that Munari and Buson
(1991) classified as an F0~V star with $B$=13.59 and {\sl B--V}=+0.45. Our
photometry gives different values, $B$=13.18 and {\sl B--V}=+0.83, with a
large scatter of 0.25 mag between three different measurements (compared to
the few millimag for nearby stars of similar brightness). All this suggests
that the optical companion is itself a variable star, and this complicates
the interpretation of photometry made with moderate or short focus
telescopes that are not able to separate QW~Sge from the close optical
companion (as it is the case for most of the archive photographic plates).
It seems relevant to observe QW Sge with enough spatial resolution to avoid
contamination from the nearby companion and to characterize the type and
amplitude of variability of the latter. If the companion should turn out to
be a moderate-amplitude variable and/or of a predictable type (like an
eclipsing system), it would be possible to remove its contribution from the
photographic photometry collected on QW Sge over the last century. No
orbital period has been determined for QW Sge.

\underline{\sl LT Del}. A large reflection effect ($\bigtriangleup U$=1.6,
$\bigtriangleup B$=0.5 and $\bigtriangleup V$= 0.2 mag) following the
ephemeris
\begin{equation}
T(min) = 2445910 (\pm 5) \ + \ 478.5 (\pm 2) \times E
\end{equation}
has been discovered by Arkhipova et al. (1995), who lists $B$ = 14.4 and
{\sl B--V}=+1.3 mag as mean values for the quiescence. MHZ report $B$=14.3
and {\sl B--V}=+1.3. The only recorded outburst of LT~Del has been
discovered in the summer of 1994 by Passuello et al. (1994), when the star
rose to $B$=12.8 and {\sl B--V}=+0.5. The star has returned to quiescence by
early 1998.

\underline{\sl Hen 2-468}. The photometric properties, history and orbital
period are unknown. MHZ list $B$=16.6 and {\sl B--V}=+1.8 mag.

\underline{\sl V407 Cyg}. Discovered as Nova Cyg 1936, it was found by
Meinunger (1966) to harbor a Mira variable with one of the longest pulsation
period known and maxima following the ephemeris
\begin{equation}
T(max) = 2429710 \ + \ 745 \times E
\end{equation}
While reconstructing the historical lightcurve, Munari et al. (1990)
discovered that brightness at the maximum of the pulsation cycle is strongly
modulated by a sinusoidal variation with a possible period around 43 years
and extrema at $B_{max}=13.3$ and $B_{max}=17.0$. The 43 year periodicity
was interpreted as the orbital period of the system. V407~Cyg was discovered
again in outburst in the summer of 1994 by Munari et al. (1994), when it
rose to $B=14.0$ and $B-V$=+1.0 (same value as in the 1936 outburst) at a
time when contemporaneous infrared photometry confirmed that the Mira was at
a minimum in its pulsation cycle. According to VSNET databank the last
minimum of the Mira has been in the very early 1998 ($V >$ 16.0) and the
last maximum in June 1999 ($V \sim$11.2). The lightcurve is however quite
complicated with humps as large as one magnitude superimposed to the much
smoother lightcurve of the Mira. The humps are possibly connected with the
current enhanced activity phase of the white dwarf and would deserve close
monitoring over the whole UBV(RI)$_{\rm C}$ range. MHZ list $B$=13.2 and
{\sl B-V}=+1.5.

\underline{\sl V627 Cas}. Originally classified among the T~Tau pre-main
sequence variables, its symbiotic star nature was discovered by Kolotilov
(1988). Kolotilov et al. (1996) has summarized the optical and infrared
photometric properties of V627 Cas. The cool component seems to be a M2
supergiant in a post-AGB phase, pulsating with a 466 day period. The hot
component presents flickering activity superimposed onto several different
types of variability, including a secular dimming by $\bigtriangleup B$=2.0
mag in 60 years. This is a peculiar type of symbiotic star which needs
more effort to be better characterized from a photometric point of view. To
null the effect of flickering, the star should be observed more times per
night and over a few consecutive nights in all the UBV(RI)$_{\rm C}$ bands.

\underline{\sl StH$\alpha$ 32}. Its symbiotic star nature has been
discovered by Downes and Keyes (1988). The photometric properties are
unknown. MHZ list $B$=14.2 and {\sl B-V}=+1.4.

\end{document}